\documentstyle[preprint,epsf]{jpsj} 

\newcommand{\lsim}
 {\ \raise.35ex\hbox{$<$}\kern-0.75em\lower.5ex\hbox{$\sim$}\ }
\def\runtitle{
Theory of Spin Polarized Tunneling in Superconducting  Sr$_{2}$RuO$_{4}$.}
\def\runauthor{Nobukatsu {\sc Yoshida}, Yukio {\sc Tanaka}, Junichiro
{\sc Inoue} 
and Satoshi {\sc Kashiwaya}}

\title
{Theory of Spin Polarized Tunneling in Superconducting Sr$_{2}$RuO$_{4}$}

\author
{
Nobukatsu {\sc Yoshida}, Yukio {\sc Tanaka}, Junichiro {\sc Inoue} 
and Satoshi {\sc Kashiwaya}$^{1}$}

\inst
{
Department of Applied Physics, Nagoya University, Nagoya 464-8603,
Japan\\
$^{1}$ Electrotechnical Laboratory, Tsukuba,
Ibaraki 305-8568, Japan}

\recdate
{December 28, 1998}

\abst
{
A theory of tunneling conductance in 
ferromagnetic-metal/insulator/triplet-superconductor 
junctions is presented for unitary and non-unitary 
spin triplet 
pairing states which are promising candidates for the 
superconducting pairing 
symmetry of Sr$_{2}$RuO$_{4}$. 
As the magnitude of the  exchange interaction in the ferromagnetic metal 
is increased, the conductance for the unitary pairing 
state below the energy gap is reduced in contrast to that for 
the non-unitary pairing state. 
This is due to the fact that the retro-reflectivity in 
the Andreev reflection for unitary pairing state is broken 
due to the influence from the exchange interaction.}

\kword
{
triplet superconductor, unitary pair potential, non-unitary
pair potential, ferromagnetic metal,
Andreev reflection
}

\begin{document}
\sloppy
\maketitle


Andreev reflection, which occurs at the 
interfaces between normal metals and 
superconductors \cite{Andreev}, is one of 
the most important elemental processes 
in electron transport through the 
superconducting junction. 
It is an interesting problem to clarify the expected 
for a tunneling effect in ferromagnet/insulator/superconductor 
($F/I/S$) junctions, 
since the retro-reflectivity of the Andreev reflection is 
broken due to an exchange interaction in the ferromagnets. 
For $s$-wave superconductors, these properties are well understood and 
new aspects of the Andreev reflection have been revealed \cite{Jong}. 
On the other hand, 
stimulated by the establishment of 
$d$-wave symmetry in high-$T_{C}$ superconductors, 
theories of the tunneling conductance of 
anisotropic superconductor junctions 
have been developed, 
which fully take into account the anisotropy of the 
pair potential \cite{Tanaka1,Tanaka2,Kashiwaya1,Kashiwaya2,
Kashiwaya3}. 
In anisotropic superconductors, the change in sign of the pair 
potential induces remarkable effects, i.e., 
a zero-bias conductance peak (ZBCP) \cite{Hu} in tunneling experiments 
of high $T_{C}$ superconductors, due to the formation of 
zero-energy states (ZES). 
Recently, previous theories have been extended 
to $F/I/S$ junctions with $d$-wave superconductors, \cite{Zutic,Zhu,Kashiwaya4} and 
the influences of the exchange interaction on transport properties 
have been clarified \cite{Kashiwaya4}. 
However, these theories treat only spin singlet 
superconductors. 
\par
The recent discovery of superconductivity 
in Sr$_{2}$RuO$_{4}$ \cite{Maeno1} 
has attracted much theoretical and experimental attention 
because this material is the first example of a 
noncuprate layered perovskite superconductor. 
Since this compound is isostructural to cuprate 
superconductors, the electronic properties in both the normal 
and the superconducting state are highly anisotropic. 
Several experiments indicate the existence of a large residual 
density of states of quasiparticles \cite{Maeno2}. 
Furthermore, the importance of 
ferromagnetic spin fluctuations 
in this material 
has been suggested \cite{Rice}. 
These results strongly imply that the
pairing states of Sr$_{2}$RuO$_{4}$ belong to 
two-dimensional triplet superconducting states, $i.e.$ 
$E_{u}$ symmetry \cite{Sigrist,Machida}. 
To clarify the  phase coherence of this material 
peculiar to anisotropic pairing, 
theories of tunneling conductance and Josephson 
effect in this material have been 
presented \cite{Yamashiro1,Yamashiro2,Honer1,Honer2,Yamashiro3}. 
However, no theory has been presented for transport properties of 
ferromagnet/insulator/triplet superconductor ($F/I/TS$) 
junctions where remarkable differences from those of 
singlet superconductor cases are expected. 
\par
In this paper, a formulation of the spin-polarized tunneling conductance in 
$F/I/TS$ junctions is presented by extending the theory for  $d$-wave 
superconductors \cite{Kashiwaya4}.  Although the superconducting state of 
Sr$_{2}$RuO$_{4}$ has yet to be fully identified, 
we will choose two types of triplet $p$-wave pair potentials: 
the unitary and the non-unitary pairing states
with $E_{u}$ symmetry. 
For the unitary pairing state, 
an incoming electron and the Andreev reflected hole have 
antiparallel spins from each other, as in the case of 
the singlet pairing state. 
In the unitary case, the conductance obtained below the energy gap 
of the superconductor is reduced drastically 
due to the exchange interaction. 
On the other hand, 
since the Andreev reflection in the non-unitary pairing state 
conserves spin, 
the influences of 
the exchange interaction are not so serious. 
Thus, the present results serve as a more useful guide for 
the experimental identification of 
the pairing state of Sr$_{2}$RuO$_{4}$. 
\par
%
For the calculation, 
a two-dimensional $F/I/TS$ junction 
with semi-infinite double-layered structures 
in the clean limit is assumed. 
A flat 
interface is perpendicular to the $x$-axis and is 
located at $x$=0. 
The insulator is modeled as 
a delta-functional form $H\delta(x)$, where $\delta(x)$ 
and $H$ are the delta-function and its amplitude, respectively. 
\noindent
The Fermi energy $E_{F}$ and the effective mass 
{\it m} are assumed to be equal both in the ferromagnet and 
in the superconductor, for simplicity. 
As a model of the ferromagnetic metal, we apply the Stoner model,
\cite{Jong} using the exchange potential $U(\mbox{\boldmath $x$})$ = 
$U\Theta(-x)$, where $\Theta(x)$ is the Heaviside step function. 
The  magnitude 
of Fermi momentum in the ferromagnet for up 
[down] spin is denoted as 
$k_{F,\uparrow}=\sqrt{\frac{2m}{\hbar^{2}}(E_{F} + U)}$ 
[$k_{F,\downarrow}=\sqrt{\frac{2m}{\hbar^{2}}(E_{F} - U)}$]. 
For simplicity, we assume the 
spatially constant pair potentials and neglect 
the effects of spin-orbit scattering. 
The wave functions $\Psi($x$)$ are obtained 
by solving the Bogoliubov-de Gennes (BdG) equation according to 
the quasiclassical approximations \cite{Bruder}. 
In this approximation, the effective pair potentials 
for quasiparticles in the superconductor 
are given by $\Delta(\theta_{S}) \Theta(x)$, 
where $\theta_{S}$ denotes the direction of the motions of quasiparticles 
which is measured from the normal to the interface. 
The pair potential matrix is expressed as 
\begin{equation}
 {\bf\Delta}(\theta_{S})=\left( \begin{array}{cc}
\Delta_{\uparrow\uparrow}(\theta_{S}) & \Delta_{\uparrow \downarrow}
(\theta_{S})\\[10pt]\Delta_{\downarrow \uparrow}(\theta_{S}) &
\Delta_{\downarrow \downarrow}(\theta_{S})
\end{array} \right).
\label{eqn:e1}
\end{equation}
In the calculation, we consider four 
kinds of pair potentials with $E_{u}$ symmetry. 
In the following, we will call $E_{u}(1)$ state for 
time  reversal symmetry state and $E_{u}(2)$ state for 
broken time  reversal symmetry state, respectively. 
For the unitary pairing state, these matrix elements are given by 
$\Delta_{\uparrow,\downarrow}(\theta_{S})
= \Delta_{\downarrow,\uparrow}(\theta_{S})$ 
=$\Delta_{0}$(sin$\theta_{S}$ + cos$\theta_{S}$), 
$\Delta_{\uparrow,\uparrow}(\theta_{S}) =
\Delta_{\downarrow,\downarrow}(\theta_{S})= 0$
for $E_{u}(1)$ state and 
$\Delta_{\uparrow,\downarrow}(\theta_{S}) =
\Delta_{\downarrow,\uparrow}(\theta_{S}) =
\Delta_{0}\exp(i\theta_{S})$,
$\Delta_{\uparrow,\uparrow}(\theta_{S}) =
\Delta_{\downarrow,\downarrow}(\theta_{S})= 0$
for $E_{u}(2)$ state, respectively. 
For non-unitary pairing state, these are 
$\Delta_{\uparrow,\uparrow}(\theta_{S})$ =
$\Delta_{0}$(sin$\theta_{S}$ + cos$\theta_{S}$),
$\Delta_{\downarrow,\downarrow}(\theta_{S})=
\Delta_{\uparrow,\downarrow}(\theta_{S}) = 
\Delta_{\downarrow,\uparrow}(\theta_{S}) =0$ 
for $E_{u}(1)$ state and 
$\Delta_{\uparrow,\uparrow}(\theta_{S}) 
= \Delta_{0}\exp(i\theta_{S})$,
$\Delta_{\downarrow,\downarrow}(\theta_{S}) =
\Delta_{\uparrow,\downarrow}(\theta_{S}) =
\Delta_{\downarrow,\uparrow}(\theta_{S}) = 0$ 
for  $E_{u}(2)$ state, 
respectively, following the discussions by 
Sigrist and Zhitomirsky\cite{Sigrist} and Machida et al\cite{Machida}. 
\par
There are four scattering processes for an electron 
injection from the ferromagnet 
with up spin and at angle $\theta_{F}$ with respect 
to the interface normal, as shown in Figure 1. 
These are Andreev reflection (AR) as a hole, 
normal reflection (NR) as an electron, 
transmission as an electron like quasiparticle (ELQ), and 
transmission as a hole like quasiparticle (HLQ). 
The transmitted ELQ and HLQ feel different effective 
pair potentials $\Delta_{ss'}(\theta_{S+})$ and 
$\Delta_{ss'}(\theta_{S-})$, 
with $\theta_{S+}=\theta_{S}$ and $\theta_{S-} = \pi-\theta_{S}$.
The wave vectors of ELQ and HLQ are approximated by 
$k_{S} = \mid \mbox{\boldmath $k$}_{S} \mid\approx \sqrt{\frac{2m E_{F}}
{\hbar^{2}}}$ in the framework of the quasiclassical approximation \cite{An
dreev}.
Since the translational symmetry holds for the $y$-axis direction, 
the momenta parallel to the interface are conserved at the interface, 
$k_{F,\uparrow}$sin$\theta_{F}$ = $k_{F,\downarrow}$sin$\theta_{F}^{'}$ 
=$k_{S}$sin$\theta_{S}$. In the case of a unitary state, 
the retro reflectivity of the Andreev reflection is 
broken due to the difference in the exchange interaction 
felt by a hole like quasiparticle.
Consequently,  $\theta_{F}$ is not equal to 
$\theta_{F}^{'}$. 

The wave function $\Psi(\mbox{\boldmath$x$})$ in the 
ferromagnet region for the unitary state is described by 
\par
\begin{equation}
\Psi(\mbox{\boldmath$x$})
=\left(
\begin{array}{c}
1 \\
0 \\
0 \\
0 \\
\end{array}
\right)e^{i\mbox{\boldmath$k$}_{F,\uparrow}\mbox{\boldmath$x$}}+
a_{\downarrow}\left(
\begin{array}{c}
0 \\
0 \\
0 \\
1 \\
\end{array}
\right)e^{i\mbox{\boldmath$k$}_{F,\downarrow}\mbox{\boldmath$x$}}+
b_{\uparrow}\left(
\begin{array}{c}
1 \\
0 \\
0 \\
0 \\
\end{array}
\right)e^{-i\mbox{\boldmath$k^{\prime}$}_{F,\uparrow}\mbox{\boldmath$x$}}.
\label{eqn:e2}
\end{equation}
That for the non-unitary state is given by 
\begin{equation}
\Psi(\mbox{\boldmath$x$})
=\left(
\begin{array}{c}
1 \\
0 \\
0 \\
0 \\
\end{array}
\right)e^{i\mbox{\boldmath$k$}_{F,\uparrow}\mbox{\boldmath$x$}}+
a_{\uparrow}\left(
\begin{array}{c}
0 \\
0 \\
1 \\
0 \\
\end{array}
\right)e^{i\mbox{\boldmath$k$}_{F,\uparrow}\mbox{\boldmath$x$}}+
b_{\uparrow}\left(
\begin{array}{c}
1 \\
0 \\
0 \\
0 \\
\end{array}
\right)e^{-i\mbox{\boldmath$k^{\prime}$}_{F,\uparrow}\mbox{\boldmath$x$}} .
\label{eqn:e3}
\end{equation}
The reflection probabilities of AR($a_{\uparrow(\downarrow)}$) and 
NR($b_{\uparrow(\downarrow)}$) are determined by solving 
the BdG equations under the boundary conditions. 
The reflection probabilities for down spin injection are also obtained 
in a similar way. 
The normalized tunneling conductance is expressed as\cite{Siki} 
\begin{equation}
\sigma_{T}(eV)=
\frac{ \int^{\pi/2}_{-\pi/2} d\theta_{S} \cos\theta_{S}
(\sigma_{S,\uparrow} +  \sigma_{S,\downarrow})}
{\int^{\pi/2}_{-\pi/2} d\theta_{S} \cos\theta_{S}
(\sigma_{N,\uparrow} + \sigma_{N,\downarrow})}
\label{eqn:e4}
\end{equation}
where $\sigma_{N,\uparrow[\downarrow]}$ denotes the tunneling 
conductance for the up[down] spin quasiparticle 
injection in the normal state and is given by 
\[
\sigma_{N,\uparrow}=
\frac{4\lambda_{+}}{(1+\lambda_{+})^{2}+Z_{\theta_{S}}^{2}},\hspace{10pt}
\lambda_{\pm}=\sqrt{1 \pm \frac{U}{E_{F}\cos^{2}\theta_{S}} }
\]
\[\sigma_{N,\downarrow}=
\frac{4\lambda_{-}}{(1+\lambda_{-})^{2}+Z_{\theta_{S}}^{2}}
\Theta(\theta_{C}-\mid \theta_{S}\mid)
\]
with $Z_{\theta_{S}}=\frac{Z}{\cos\theta_{S}}$ and 
$Z=\frac{2mH}{\hbar^{2} k_{F}}$. 
\par
The quantity $\sigma_{S,\uparrow[\downarrow]}$ is the tunneling 
conductance for the up[down] spin electron injection 
in the superconducting state. 
In the unitary pairing state, we note that the 
Fermi surface effect largely influences the 
reflection process. 
The retro-reflectivity of AR is broken due to the influence of 
the exchange interaction in the ferromagnet. 
In the following, we will consider the situation where 
$k_{F,\downarrow}< k_{S} < k_{F,\uparrow}$
is satisfied [Fig. 2]. 
For $|\theta_{S}| > \sqrt{\cos^{-1}(U/E_{F})}\equiv\theta_{C}$, 
the Andreev reflection does not exist 
as a propagating wave. 
This novel property in the $F/I/S$ junction is caused by 
the fact that 
the Andreev reflected hole has an antiparallel spin to that of the 
injection electron. 
\par
Based on the calculation of singlet superconductors 
\cite{Kashiwaya1,Kashiwaya2,Kashiwaya3,Kashiwaya4}, 
the tunneling conductance $\sigma_{S,\uparrow[\downarrow]}$ 
is given by 
\par
\[
\sigma_{S,\uparrow}
=
\sigma_{N,\uparrow}
\frac{1-\mid \Gamma_{+} \Gamma_{-} \mid^{2}(1-\sigma_{N,\downarrow})
+ \sigma_{N,\downarrow} \mid{\Gamma_{+}} \mid^{2} }
{\mid 1 - \Gamma_{+} \Gamma_{-}
\sqrt{1-\sigma_{N,\downarrow}} \sqrt{1-\sigma_{N,\uparrow}}
\exp[i(\varphi_{\downarrow}-\varphi_{\uparrow})] \mid^{2}}
\Theta(\theta_{C}-\mid \theta_{S}\mid)
\]
\begin{equation}
+[1-\Theta(\theta_{C}-\mid \theta_{S}\mid )]
\sigma_{N,\uparrow}
\frac{[1-\mid \Gamma_{+} \Gamma_{-} \mid^{2}]}
{\mid 1 - \Gamma_{+} \Gamma_{-}
\sqrt{1-\sigma_{N,\uparrow}}
\exp[i(\varphi_{\downarrow}-\varphi_{\uparrow}) \mid^{2}}
\end{equation}
\begin{equation}
\sigma_{S,\downarrow}
=
\sigma_{N,\downarrow}
\frac{1-\mid \Gamma_{+} \Gamma_{-} \mid^{2}(1-\sigma_{N,\uparrow})
+ \sigma_{N,\uparrow} \mid{\Gamma_{+}} \mid^{2}}
{\mid 1 - \Gamma_{+} \Gamma_{-}
\sqrt{1-\sigma_{N,\downarrow}} \sqrt{1-\sigma_{N,\uparrow}}
\exp[i(\varphi_{\uparrow}-\varphi_{\downarrow}) \mid^{2}}
\Theta(\theta_{C} -\mid\theta_{S}\mid )
\end{equation}
\[
\exp(i\varphi_{\downarrow})=
\frac{1-\lambda_{-} + iZ_{\theta_{S}}}
{\sqrt{1 - \sigma_{N,\downarrow}} (1+\lambda_{-} - iZ_{\theta_{S}}) },
\exp(-i\varphi_{\uparrow})=
\frac{1-\lambda_{+} - iZ_{\theta_{S}}}
{\sqrt{1 - \sigma_{N,\uparrow}} (1+\lambda_{+} + iZ_{\theta_{S}}) }
\]
for the unitary pairing state and
\begin{equation}
\sigma_{S,\uparrow}
=
\sigma_{N,\uparrow}
\frac{1-\mid \Gamma_{+} \Gamma_{-} \mid^{2}(1-\sigma_{N,\uparrow})
+ \sigma_{N,\uparrow} \mid{\Gamma_{+}} \mid^{2} }
{\mid 1 - \Gamma_{+} \Gamma_{-}
(1-\sigma_{N,\uparrow}) \mid^{2}}
\end{equation}
\begin{equation}
\sigma_{S,\downarrow}=\sigma_{N,\downarrow}
\end{equation}
for the non-unitary pairing sate. 
Here, $\Gamma_{\pm}$, defined for both 
unitary and non-unitary pairing states, is denoted by 
\par
\[
\Gamma_{\pm}=\frac{\Delta_{0}(\sin\theta_{S} \pm \cos\theta_{S})}
{eV+\sqrt{(eV)^{2}-
|\Delta_{0}(\sin\theta_{S} \pm \cos\theta_{S})|^{2}}},
\hspace{10pt}   E_{u}(1)
\]
\[
\Gamma_{\pm}=\pm\frac{eV-\sqrt{(eV)^{2}-|\Delta_{0}|^{2}}}
{|\Delta_{0}|} e^{-i\theta_{S}},
\hspace{40pt}   E_{u}(2)
\]
\par
\par
In the above formulations, when the ferromagnet is a 
normal metal ($i.e.$,
$U=0$), $\sigma_{T}(eV)$ in refs. 18 and 19 
are reproduced completely.
On the other hand,  the magnitude of the Fermi momentum 
becomes zero for down spin electrons for 
the half-metallic ferromagnet limit ($i.e.$, $U=E_{F}$), 
and $\sigma_{N,\downarrow}=0$. 
Then, the wave function of the reflected hole by AR becomes 
an evanescent 
wave, and hence does not contribute to the net current. 
In this case, 
since the numerator of the conductance formula in eq. (3) 
vanishes at $eV$=0, 
the ZBCP disappears both for the $Eu(1)$ and $Eu(2)$ pair potentials. 
In particular for the $Eu(2)$ pair potential, 
since the $\mid\Gamma_{+}\Gamma_{-}\mid^{2}$ = 1 is satisfied 
independent of $\theta_{S}$, 
for $eV < \Delta_{0}$,  the numerator is zero and the 
resulting conductance $\sigma_{S,\uparrow}(eV)=0$. 
%
\par
Figure 3 shows the calculated conductance spectra of the $Eu(1)$ pair 
potential for both unitary and non-unitary paring states 
with various $X=U/E_{F}$ for $Z=0$. 
The results in refs. 18 and 19 are reproduced for $X=0$. 
In the unitary pairing state [see Fig. 3(a)], 
since the probability of the Andreev 
reflection is suppressed due to finite $U$, 
the tunneling conductance inside the gap ($eV<\Delta_{0}$) 
is drastically reduced with the increase of $X$, similar to 
the case of $d$-wave superconductor\cite{Kashiwaya4}. 
On the other hand for the non-unitary 
pairing state [see Fig. 3(b)], 
the line shape of $\sigma_{T}(eV)$ is 
insensitive with increasing $X$. 
Figure 4 shows the $\sigma_{T}(eV)$ of the $Eu(2)$ pair potential 
in the high barrier case ($Z=5$) for various $X$. 
In this case, $\sigma_{T}(eV)$ has a ZBCP 
due to the sign change of the pair potential. 
The height of this peak for the unitary pairing state 
[see Fig. 4(a)] is 
reduced with increasing $X$ in contrast to that for the non-unitary 
case [see Fig. 4(b)].
Since $\sigma_{S,\downarrow}=0$ for $|\theta_{S}| > \theta_{C}$
for the unitary pairing state with $eV<\Delta_{0}$,
$\sigma_{T}(eV)$ disappears with the increase of $X$. 
When the magnetization axis of 
the ferromagnet is antiparallel to the spin axis of 
the non-unitary pair potential, the height of the ZBCP 
is reduced and $\sigma_{T}(eV)$ converges to unity with 
increasing $X$ [see Fig. 4(c)]. 
Similar to the $d$-wave superconductor case\cite{Kashiwaya4},
we can estimate the magnitude of the spin polarization 
using the height of the ZBCP. 
\par
\noindent
\par
In conclusion, 
we have studied the properties of tunneling 
conductance spectra $\sigma_{T}(eV)$ in $F/I/TS$ junctions. 
In the case of a unitary pair potential with $E_{u}$ symmetry 
where the Cooper pair is formed between up and down spins, 
the height of the ZBCP is reduced drastically 
as the magnitude of the exchange interaction is increased. 
This is due to the breakdown of the retro-reflectivity of the 
Andreev reflection. 
For the non-unitary case, 
the conductance depends on 
the direction of the magnetization axis of the ferromagnet. 
When the magnetization axis is parallel (antiparallel) to the 
spin axis of the non-unitary pair potential, the height of the ZBCP 
is enhanced (suppressed) with the increase of the 
exchange interaction. 
Since a Cooper pair is 
formed between quasiparticles with equal spins, 
the exchange interaction does not significantly influence on 
the Andreev reflection. 
Based on these properties, we expect that the 
symmetry of the pair potential in 
Sr$_{2}$RuO$_{4}$ and the magnitude of the exchange 
interaction can be identified in future experiments 
by the presence of $F/I/TS$ junction . 
\par
We would like to thank M. Yamashiro and H. Itoh for useful 
discussions. This work has been partially supported by the Core 
Research for Evolutional Science and Technology 
(CREST) of the Japan Science and Technology 
Corporation (JST).
%

\newpage
\begin{figure}
\vspace{20pt}
\begin{center}
\leavevmode
 \epsfxsize=80mm
 \epsfbox{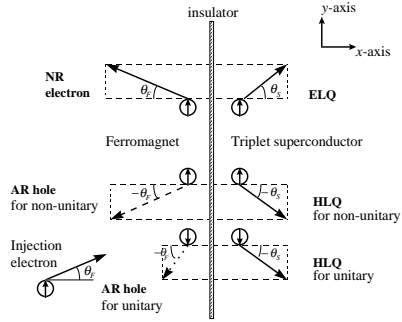}
\end{center}
\caption{Schematic illustration of the scattering
process of the quasiparticle at the interface of the
$F/I/TS$ junction. For the Andreev reflected hole with down spin,
the retro-reflectivity is broken due to the exchange interaction. }
\label{fig:f1}
\end{figure}
%
\begin{figure}
\vspace{20pt}
\begin{center}
\leavevmode
 \epsfxsize=80mm
 \epsfbox{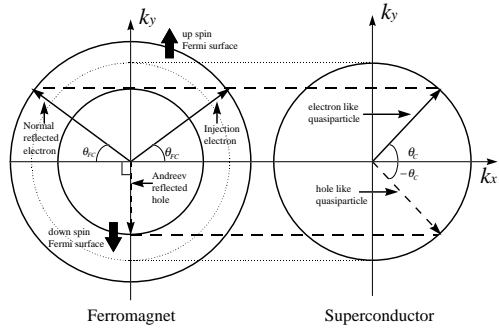}
\end{center}
\caption{Fermi surface effect for Andreev reflection for 
the unitary pairing state. 
$\theta_{FC}$ and $\theta_{C}$ 
are the angles in the ferromagnet 
and in the triplet superconductor, respectively. 
For $\theta_{FC}<|\theta_{F}|$, $i.e.$, 
$\theta_{C}<|\theta_{S}|$, 
the Andreev reflection does not exist as a propagating wave. }
\label{fig:f2}
\end{figure}
%
\begin{figure}
\vspace{20pt}
\begin{center}
\leavevmode
 \epsfxsize=80mm
 \epsfbox{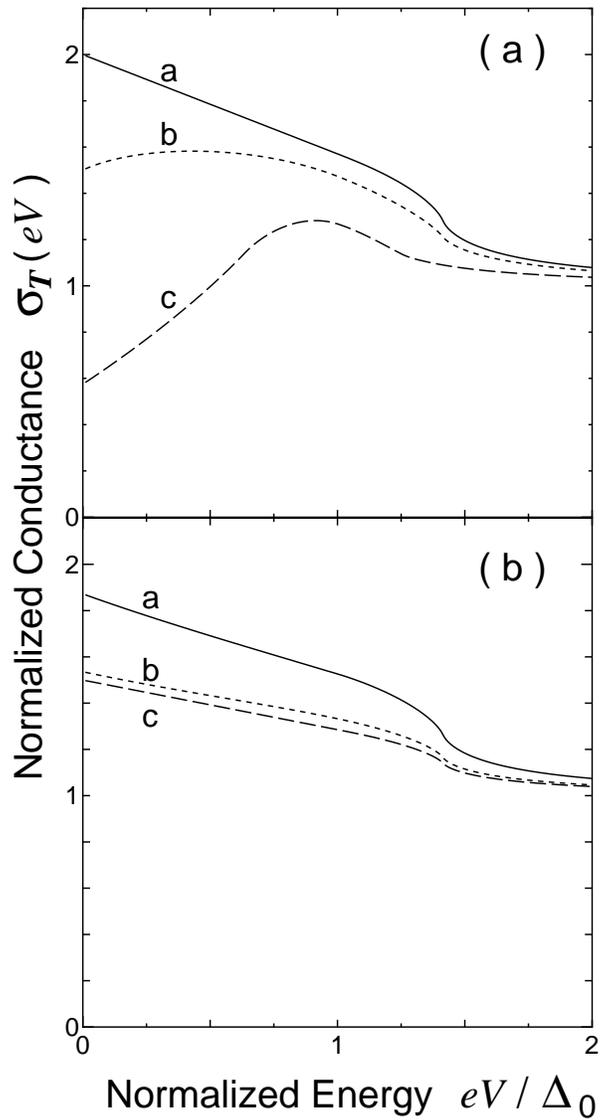}
\end{center}
\caption{Normalized conductance spectra for
$E_{u}(1)$ symmetry as a function of $X$ with $Z=0$.
Here (a) and (b) are results for the unitary pairing state
and the non-unitary pairing state, respectively.
$X$ is a: $X=0.0$, b: $X=0.5$ and c: $X=0.9$ for the unitary pairing state
 and $X=0.999$ for the non-unitary pairing state. }
\label{fig:f3}
\end{figure}
%
\begin{figure}
\vspace{20pt}
\begin{center}
\leavevmode
 \epsfxsize=80mm
 \epsfbox{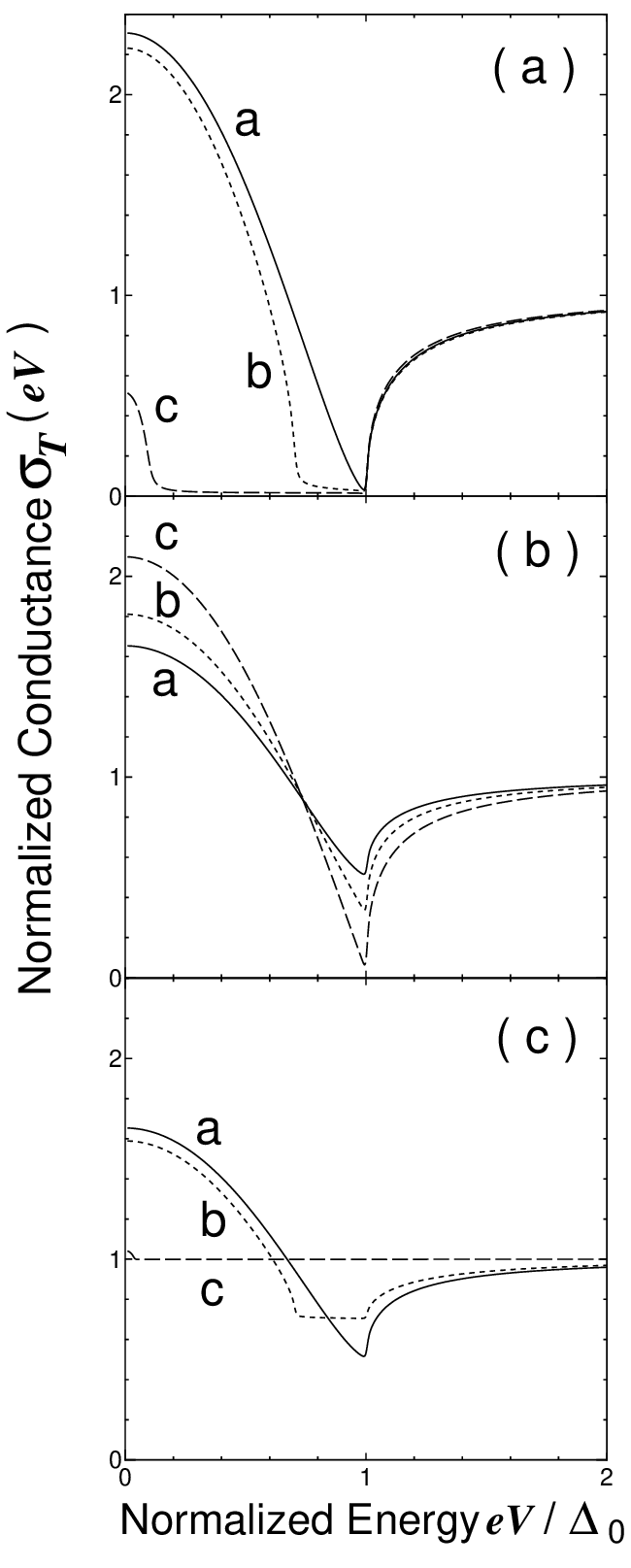}
\end{center}
\caption{Normalized conductance spectra for
$E_{u}(2)$ symmetry as a function of $X$ with $Z=5$.
Here, (a) and (b) are results for the unitary
pairing state and the non-unitary pairing state,
respectively. 
(c) is the result of 
the magnetization axis of ferromagnet antiparallel 
to the spin axis of the non-unitary pair potential. 
$X$ is a: $X=0.0$, b: $X=0.5$ and c: $X=0.999$. }
\label{fig:f4}
\end{figure}

\begin{thebibliography}{99}
\bibitem{Siki}
In contrast to ref. 7, we neglect the polarization factor in eq. (4). 
This point is not essential because the inclusion of the factor depends on
the model.
\end{thebibliography}
\end{document}